\documentclass{PoS}
\usepackage{subcaption}

\def\slash#1{\mbox{$\not \!\! #1$}}
\def\Dslash{{\slash {\cal D}}}

\def\lvec#1{\setbox0=\hbox{$#1$}
    \setbox1=\hbox{$\scriptstyle\leftarrow$}
    #1\kern-\wd0\smash{
    \raise\ht0\hbox{$\raise1pt\hbox{$\scriptstyle\leftarrow$}$}}
    \kern-\wd1\kern\wd0}
\def\rvec#1{\setbox0=\hbox{$#1$}
    \setbox1=\hbox{$\scriptstyle\rightarrow$}
    #1\kern-\wd0\smash{
    \raise\ht0\hbox{$\raise1pt\hbox{$\scriptstyle\rightarrow$}$}}
    \kern-\wd1\kern\wd0}

\def\diracstar#1#2{
    \setbox0=\hbox{$\gamma$}\setbox1=\hbox{$\gamma_{#1}$}
    \gamma_{#1}\kern-\wd1\kern\wd0
    \smash{\raise4.5pt\hbox{$\scriptstyle#2$}}}

\def\tr{\,\hbox{tr}\,}

\newcommand{\beq}{\begin{equation}}
\newcommand{\eeq}{\end{equation}}
\newcommand{\beqn}{\begin{eqnarray}}
\newcommand{\eeqn}{\end{eqnarray}}
\newcommand{\nn}{\nonumber}

\def\com#1{\ifcomment\textcolor{red}{[#1]}\fi}
\graphicspath{{figures/}}

\newif\ifcomment
 \commentfalse
\title{Check of a new non-perturbative mechanism for elementary fermion mass generation}

\ShortTitle{Non-perturbative elementary fermion mass generation}

\author{
S.\ Capitani$^{a)}$, G.M.\ de Divitiis$^{b)}$, P.\ Dimopoulos$^{c)}$, R.\ Frezzotti$^{b)}$,
 \speaker{M.\ Garofalo}$^{d)}$, B.\ Knippschild$^{e)}$,
 B.\ Kostrzewa$^{e)}$, K.\ Ottnad$^{e)}$, G.C.\ Rossi$^{b)c)}$, M.\ Schr\"ock$^{f)}$, C.\ Urbach$^{e)}$

\\
$^{a)}$ Johann Wolfgang Goethe-Universit\"at Frankfurt am Main
Institut f\"ur Theoretische Physik
Max-von-Laue-Stra\ss e 1
D-60438 Frankfurt am Main
Germany
\\
$^{b)}$ 
  Dipartimento di Fisica, Universit\`a di  Roma
  ``{\it Tor Vergata}'' and INFN, Sezione di Roma 2,
     Via della Ricerca Scientifica - 00133 Rome, Italy\\
 $^{c)}$ Centro Fermi - Museo Storico della Fisica e Centro Studi e Ricerche Enrico Fermi, Compendio del Viminale, Piazza del Viminiale 1, I-00184, Rome, Italy\\
  $^{d)}$
Higgs Centre for Theoretical Physics, School of Physics and Astronomy,
    The University of Edinburgh, Edinburgh EH9 3JZ, Scotland, UK  \\
 $^{e)}$ 
 Institut f\"ur Strahlen-und Kernphysik (Theorie), Nussallee 14-16 Bethe Center for Theoretical Physics, Nussallee 12 Universit\"at Bonn, D-53115 Bonn, Germany\\
  $^{f)}$  Istituto Nazionale di Fisica Nucleare (INFN), Sezione di Roma Tre, 00146 Rome, Italy\\
        E-mail: \email{marcogarofalo31@gmail.com}}

\abstract{We consider a field theoretical model where a SU(2) fermion doublet, subjected to non-Abelian gauge interactions, is also coupled to a complex scalar field doublet via a Yukawa and an irrelevant Wilson-like term. Despite the presence of these two chiral breaking operators in the Lagrangian, an exact symmetry acting on fermions and scalars prevents perturbative mass corrections. In the phase where fermions are massless (Wigner phase) the Yukawa coupling can be tuned to a critical value at which chiral transformations acting on fermions only become a symmetry of the theory (up to cutoff effects). In the Nambu-Goldstone phase of the critical theory a fermion mass term of dynamical origin is expected to arise in the Ward identities of the purely fermionic chiral transformations. Such a non-perturbative mechanism of dynamical mass generation can provide a ``natural'' (\`a la 't Hooft) alternative to the Higgs mechanism adopted in the Standard Model.
Here we lay down the theoretical framework necessary to demonstrate
the existence of this mechanism by means of lattice simulations.

}

\FullConference{34th annual International Symposium on Lattice Field Theory\\
		24-30 July 2016\\
		University of Southampton, UK}

\begin{document}

\section{The mechanism in a simple model}\label{The_model}
In~\cite{FR_mechanism} a new non-perturbative (NP) mechanism for the elementary particle mass generation was conjectured. Existence and viability of this phenomenon can be 
tested in the toy model described by the Lagrangian
\begin{eqnarray}
\hspace{-1.6cm}&&{\cal L}_{\rm{toy}}(Q,A,\Phi)= {\cal L}_{kin}(Q,A,\Phi)+{\cal V}(\Phi)
+{\cal L}_{Wil}(Q,A,\Phi) + {\cal L}_{Yuk}(Q,\Phi) \, ,\label{SULL} \\
\hspace{-1.6cm}&&\quad{\cal L}_{kin}(Q,A,\Phi)= \frac{1}{4}(F\cdot F)+\bar Q_L\Dslash Q_L+\bar Q_R\Dslash \,Q_R+\frac{1}{2}{\tr}\big{[}\partial_\mu\Phi^\dagger\partial_\mu\Phi\big{]}\label{LKIN}\\
\hspace{-1.6cm}&&\quad{\cal V}(\Phi)= \frac{\mu_0^2}{2}{\tr}\big{[}\Phi^\dagger\Phi\big{]}+\frac{\lambda_0}{4}\big{(}{\tr}\big{[}\Phi^\dagger\Phi\big{]}\big{)}^2\label{LPHI}\\
\hspace{-1.6cm}&&\quad{\cal L}_{Wil}(Q,A,\Phi)= \frac{b^2}{2}\rho\,\big{(}\bar Q_L{\overleftarrow{\cal D}}_\mu\Phi {\cal D}_\mu Q_R+\bar Q_R \overleftarrow{\cal D}_\mu \Phi^\dagger {\cal D}_\mu Q_L\big{)}
\label{LWIL} \\
\hspace{-1.6cm}&&\quad{\cal L}_{Yuk}(Q,\Phi)=\
  \eta\,\big{(} \bar Q_L\Phi Q_R+\bar Q_R \Phi^\dagger Q_L\big{)}
\label{LYUK} \, ,
\end{eqnarray}
where $b^{-1}=\Lambda_{UV}$ is the UV-cutoff.
The Lagrangian~(\ref{SULL}) describes a non-Abelian gauge model where an SU(2) doublet of strongly interacting fermions is coupled to a complex scalar field via Wilson-like (eq.~(\ref{LWIL})) and Yukawa (eq.~(\ref{LYUK})) terms. For short we have used a compact SU(2)-like notation where $Q_L=(u_L\,\,d_L)^T$ and $Q_R=(u_R\,\,d_R)^T$ are fermion iso-doublets and $\Phi$ is a $2\times2$ matrix with $\Phi=(\phi,-i\tau^2 \phi^*)$ and $\phi$ an iso-doublet of complex scalar fields.
\\
The term ${\cal V}(\Phi)$ in eq.~(\ref{LPHI}) is the standard quartic scalar potential where the (bare) parameters $\lambda_0$ and $\mu_0^2$ control the self-interaction and the mass of the scalar field. In the equations above we have introduced the covariant derivatives
\beq
{\cal D}_\mu=\partial_\mu -ig_s \lambda^a A_\mu^a \, , \qquad
\overleftarrow{\cal D}_\mu =\overleftarrow{\partial}_\mu +ig_s \lambda^a A_\mu^a \, ,\label{COVG}
\eeq
where $A_\mu^a$ is the gluon field ($a=1,2,\dots, N_c^2-1$) with field strength $F_{\mu\nu}^{a}$.
A crucial r\^ole in the model is played by the $d=4$ Yukawa term ${\cal L}_{Yuk}$ and the Wilson-like $d=6$ operator  ${\cal L}_{Wil}$. For dimensional reasons the latter enters the Lagrangian multiplied by $b^2$.

Besides Lorentz, gauge and $C$, $P$, $T$, $CPF_2$ symmetries (see Appendix B of~\cite{FR_mechanism}), ${\cal L}_{\rm toy}$ is invariant under the following (global) transformations $\chi_L$ and $\chi_R$
\beqn
&&\bullet\,\chi_{L}:\quad \tilde\chi_{L}\otimes (\Phi\to\Omega_L\Phi) 
\quad\quad\bullet\,\chi_{R}:\quad \tilde\chi_{R}\otimes (\Phi\to\Phi\Omega_R^\dagger) 
\label{CHIL}\\
\hspace{-2.cm}&&\tilde\chi_{L/R} : \left \{\begin{array}{l}     
Q_{L/R}\rightarrow\Omega_{L/R} Q_{L/R}  \\
\hspace{4cm}\Omega_{L/R}\in {\mbox{SU}}(2)_{L/R}\\
\bar Q_{L/R}\rightarrow \bar Q_{L/R}\Omega_{L/R}^\dagger \\ 
\end{array}\right . \label{GTWT}
\eeqn
The model~(\ref{SULL}) is power-counting renormalizable (as LQCD is) with counter-terms constrained by the exact symmetries of the Lagrangian. In particular, owing to the presence of the scalar field and the related exact $\chi_L \otimes \chi_R$ symmetry, no power divergent fermion mass terms can be generated. 

\subsection{Fermionic chiral symmetry enhancement}

For generic values of the parameters $(\rho,\eta) \neq (0,0)$ , ${\cal L}_{\rm{toy}}$ is not invariant under the chiral transformations $\tilde\chi_L$  and $\tilde\chi_R$ 
(eq.~(\ref{GTWT})). We are interested in the case where fermionic chiral symmetries 
are not exact as the breaking terms can polarize the vacuum under dynamical symmetry
breaking due to strong interactions. To study possible enhancement of $\tilde\chi_L$
symmetry (by parity the same will hold also for $\tilde\chi_R$) we consider the
(bare) $\tilde\chi_L$ WTIs, v.i.z.\
\beqn
\hspace{-0.7cm}&&\partial_\mu \langle \tilde J^{L\, i}_\mu(x) \,\hat{\cal O}(0)\rangle = \langle \tilde\Delta_{L}^i \hat{\cal O}(0)\rangle\delta(x) -
\eta \,\langle \big{(} \bar Q_L\frac{\tau^i}{2}\Phi Q_R-\bar Q_R\Phi^\dagger\frac{\tau^i}{2}Q_L \big{)}(x)\,\hat{\cal O}(0)\rangle \!+\nn\\
\hspace{-1.2cm}&&\phantom{\partial_\mu J^{L\, i}_\mu}-\frac{b^2}{2}\rho\,\langle\Big{(} \bar Q_L\overleftarrow {\cal D}_\mu\frac{\tau^i}{2}\Phi{\cal D}_\mu Q_R-\bar Q_R\overleftarrow {\cal D}_\mu \Phi^\dagger\frac{\tau^i}{2}{\cal D}_\mu Q_L\Big{)}(x)\,\hat{\cal O}(0)\rangle\, ,\label{CTLTI} 
\eeqn
where $\tilde\Delta_{L}^i\hat{\cal O}(0)$ is the variations of $\hat{\cal O}(0)$ under $\tilde\chi_L$ and the non-conserved currents associated are 
\beq
\tilde J_\mu^{L\,i}= \bar Q_L\gamma_\mu\frac{\tau^i}{2}Q_L -\frac{b^2}{2}\rho\Big{(}\bar Q_L\frac{\tau^i}{2}\Phi {\cal D}_\mu Q_R - \bar Q_R\overleftarrow {\cal D}_\mu\Phi^\dagger\frac{\tau^i}{2} Q_L\Big{)}\, .
\label{JCLT}
\eeq
Under renormalization the $d = 6$ operator 
$ \; O_{6}^{L\, i} = \frac{1}{2} \rho
\Big{[}\bar Q_L\overleftarrow {\cal D}_\mu\frac{\tau^i}{2}\Phi{\cal D}_\mu Q_R
-{\mbox {h.c.}}\Big{]} \; $ 
mixes with two $d = 4$ operators, plus a set of six-dimensional ones that we globally denote by $[O_{6}^{L\, i}]_{sub}$~\footnote{We do not need to resolve the mixing among the different $d=6$ operators, as they only yield negligible O($b^2$) effects. To simplify the mixing pattern (\ref{O6L-MIX}) we have used $\partial_\mu J^{L,i}_{\mu}=0$, where $J_\mu^{L,i}$ is the Noether current associated with the exact symmetry $\chi_L$ (sec.~\ref{The_model})}, 
v.i.z.\
\beqn
\hspace{-1.2cm}&& O_{6}^{L\, i} = 
\Big{[} O_{6}^{L\, i}  \Big{]}_{sub} +
\frac{Z_{\tilde{J}}-1}{b^{2}}\partial_\mu\tilde J^{L\, i}_\mu
-\frac{\bar\eta}{b^{2}}\Big{[}\bar Q_L\frac{\tau^i}{2}\Phi Q_R
-{\mbox {h.c.}}\Big{]} + \ldots
\label{O6L-MIX}
\eeqn
where $Z_{\tilde{J}}$ and $\bar\eta$ are functions of the dimensionless bare parameters entering~(\ref{SULL}) and hence depend on the subtracted scalar squared mass $\mu_{sub}^2=\mu_0^2-b^2\tau$  through the combination $b^2\mu_{sub}^2$ that is a negligible $O(b^2 )$ quantity \cite{FR_mechanism}. Thus we write 
$Z_{\tilde{J}}=Z_{\tilde{J}}(\eta;g^2_s,\rho,\lambda_0)$ and $\bar\eta=\bar\eta(\eta;g^2_s,\rho,\lambda_0)$.\com{???} Ellipses in the r.h.s.\ of eqs.~(\ref{O6L-MIX}) denote possible NP contributions to operator mixing, the possible occurrence of which will be discussed below. Plugging  (\ref{O6L-MIX}) in to (\ref{JCLT}) we get 
\begin{equation}
 \partial_\mu \langle Z_{\tilde{J}}\tilde J^{L,i}_\mu(x) \,\hat {\cal O}(0)\rangle\! = \!\langle \tilde\Delta_{L}^i \hat {\cal O}(0)\rangle\delta(x) -
({\eta- \overline\eta}) \,\langle {O_{Yuk}^{L,i}}(x)\,\hat {\cal O}(0)\rangle +{\ldots}+{\mbox{O}(b^2)}\nn. \label{ren_WTI}
\end{equation}
Setting $\eta=\eta_{cr}(g^2_s,\rho,\lambda_0)$ such that $\eta_{cr}(g^2_s,\rho,\lambda_0)-\bar\eta(\eta_{cr}; g^2_s,\rho,\lambda_0)=0$ the WTI become
\begin{equation}
 \partial_\mu \langle Z_{\tilde{J}}\tilde J^{L,i}_\mu(x) \,\hat {\cal O}(0)\rangle\! = \!\langle \tilde\Delta_{L}^i \hat {\cal O}(0)\rangle\delta(x) 
 +{\ldots}+{\mbox{O}(b^2)}\label{SYMCH} \, ,
\end{equation}
implying restoration of the fermionic $\tilde\chi_L \otimes \tilde\chi_R$ symmetries up to O($b^2$)  UV cutoff effects.

\subsection{Mass generation mechanism in the critical model} 
The physics of the model~(\ref{SULL}) at the critical value $\eta_{cr}$ crucially depends on whether the parameter $\mu^2_0$ is such that ${\cal V}(\Phi)$ has a unique minimum (Wigner phase of the $\chi_L$ symmetry, $\mu_{sub}^2 > 0$) or whether ${\cal V}(\Phi)$ develops the typical ``mexican hat'' shape (Nambu--Goldstone phase $\mu_{sub}^2 < 0$). 
In the Wigner phase no NP terms (i.e.\ ellipses) are expected to occur in the mixing pattern of eq.~(\ref{O6L-MIX}) and the transformations $\tilde \chi_L$ leads to eq.~(\ref{SYMCH}) without the ellipses.


In the Nambu-Goldstone phase a non-perturbative term is expected/conjectured\cite{FR_mechanism} to appear in the mixing pattern of eqs.~(\ref{O6L-MIX}) leading to a WTI of the form
\begin{equation}
  \partial_\mu \langle Z_{\tilde J}\tilde J^{L,i}_\mu(x) \,\hat {\cal O}(0)\rangle_{\eta_{cr}} = \langle \tilde\Delta_{L}^i \hat {\cal O}(0)\rangle_{\eta_{cr}}\delta(x)+\langle {C_1\Lambda_s[ \overline Q_L \frac{\tau^i}{2}{\cal U} Q_R+\mbox{h.c.}]} \hat {\cal O}(0)\rangle +{\mbox O}(b^2)\nn
\end{equation}
where
\beq 
{\cal U} = \frac{\Phi}{\sqrt{\Phi^\dagger \Phi}}=\frac{v+\sigma+i\overrightarrow{\tau}\overrightarrow{\pi}}{\sqrt{(v+\sigma)^2+\overrightarrow{\pi}\overrightarrow{\pi}}}\, .\label{U}
\eeq 
${\cal U}$ is a dimensionless non-analytic function of $\Phi$ that has the same transformation properties as the latter under $\chi_L \times \chi_R$ and is well defined only if $\langle\Phi\rangle=v\neq 0$.
In the local effective action $\Gamma^{NG}_{loc}$ of the theory the term $C_1\Lambda_s[ \overline Q_L {\cal U} Q_R+\mbox{hc}]$ plays the role of a mass term. 
It does not stem from the Yukawa term and, interestingly, can give a natural  (in the sense of 't~Hooft~\cite{t_hooft})  understanding of the fermion mass hierarchy problem (see discussion in \cite{FR_mechanism}).

\begin{center}
\begin{figure}[!hpbt]
\includegraphics[scale=0.26, trim=-100 0 0 0]{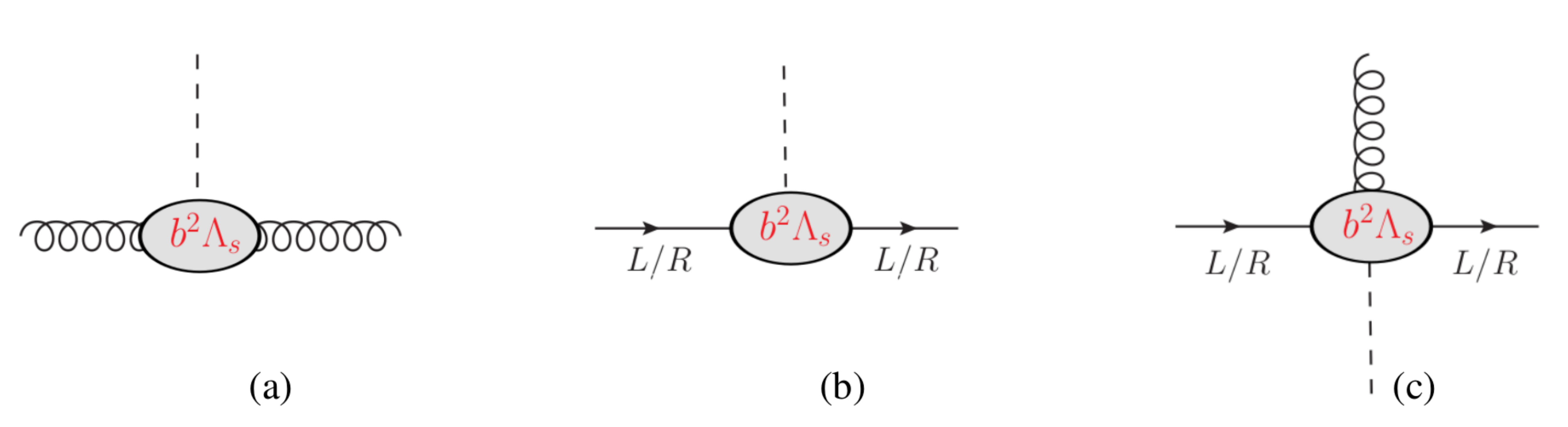}
\caption{Some of the NP O($b^2\Lambda_s\alpha_s^2$) effective vertices that are conjectured to arise~\cite{FR_mechanism} in the Nambu-Goldstone phase of the model.}
\label{fig:vertices}
\end{figure}
\end{center}

\begin{center}
\begin{figure}[!hpbt]
\vspace{-1.cm}
 \includegraphics[width=\textwidth,trim= 0 0 0 0]{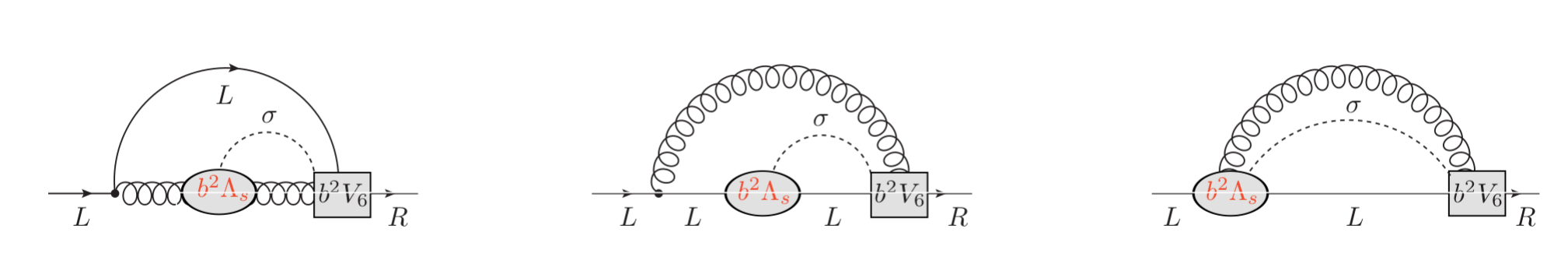}
\caption{Typical lowest order self-energy "diagrams" giving rise to dynamically generated
quark mass terms. The grey box represents the insertion of the Wilson-like vertex stemming from ${\cal L}_{Will}$. The dotted line represents the propagation of a scalar particle. The $b^{-4}$ loop divergency is cancelled by the two vertices O$(b^2)$ giving rise to a finite result.}
\label{fig:self_energy}
\end{figure}
\end{center}

An idea of how the mechanism works can be obtained from a perturbative expansion where Feynman diagrams are evaluated with the Lagrangian (\ref{SULL}) augmented 
by few extra term representing the expected $O(b^2)$ NP effective vertices \cite{FR_mechanism}, as those shown in fig.~\ref{fig:vertices}
These vertices can be inserted together with $O(b^2)$ vertices coming from the term (\ref{LWIL}) in diagrams like the ones depicted in fig.~\ref{fig:self_energy}, giving rise to finite self-energy contributions.\\
It is worth noticing that if the mechanism we have conjectured really exist it will generate a NP mass therm for the fermions even in the quenched approximation where the vertices (b) and (c) of fig.~\ref{fig:vertices}, and thus the two rightmost diagrams of fig.~\ref{fig:self_energy}, are still present.
\vspace{-0.2cm}

\vspace{-1.cm}
\section{Lattice quenched study of ${\cal L}_{toy}$: regularization and renormalization }

Numerical simulations of lattice models with gauge, fermions and scalars are not common and technically 
challenging\footnote{To our knowledge this is the first numerical study of a model with fermions, scalars and non-Abelian gauge fields in strong interaction regime.}. 
In this first numerical study of the model~(\ref{SULL}) we can limit ourselves to a 
{\em quenched-fermion} simulation of the lattice regularized action
\begin{eqnarray}
&&{ { S}_{{ lat}}=b^4\sum_x\Big{\{}{\cal L}_{kin}^{YM}[U]+{\cal L}_{kin}^{sca}(\Phi)+{\cal V}(\Phi)}{+ \overline  \Psi  { D_{lat}[U,\Phi]}  \Psi\Big{\}}}  
\label{L_lat}\\ 
 &&{\cal L}_{kin}^{YM}[U]\,\,{\mbox{: SU($3$) plaquette action}}  
\\ 
 &&{\cal L}_{kin}^{sca}(\Phi)+{\cal V}(\Phi)=\frac{1}{2}{\tr}[\Phi^\dagger(-\partial_\mu^*\partial_\mu)\Phi]+\frac{\mu_0^2}{2}{\tr}\big{[}\Phi^\dagger\Phi\big{]}+\frac{\lambda_0}{4}\big{(}{\tr}\big{[}\Phi^\dagger\Phi\big{]}\big{)}^2 \,,
\end{eqnarray}
where we have set $\Phi = \varphi_01\!\!1+i \varphi_j\tau^j$ 
\begin{eqnarray}
 (D_{lat}[U,\Phi]&&\!\! \Psi)(x)=
\gamma_\mu \widetilde\nabla_\mu  \Psi(x) + \eta F(x)  \Psi(x)
-b^2\rho \frac{1}{2}F(x)
\widetilde\nabla_\mu \widetilde\nabla_\mu  \Psi(x)
\\ 
&& - b^2\rho \frac{1}{4} \Big{[} (\partial_\mu F)(x) U_\mu(x) \widetilde\nabla_\mu  \Psi(x+\hat\mu) + (\partial_\mu^* F)(x) U_\mu^\dagger(x-\hat\mu) \widetilde\nabla_\mu  \Psi(x-\hat\mu) \Big{]} \,,
\end{eqnarray}
 with $\; F(x) \equiv [\varphi_0 1\!\!1
+i\gamma_5\tau^j\varphi_j](x) $ and the lattice derivatives defined as
\begin{eqnarray}
&&\nabla_\mu f(x)\equiv \frac{1}{b}(U_\mu(x)f(x+\hat\mu)-f(x))\quad\quad\nabla_\mu^* f(x)\equiv \frac{1}{b}(f(x)-U_\mu^\dagger(x-\hat\mu)^f(x-\hat\mu))\label{forward_back_derivative}\\
&&\widetilde\nabla_\mu f(x)\equiv \frac{1}{2}(\nabla_\mu+\nabla_\mu^*)F(x)
\end{eqnarray}
The Lagrangian (\ref{L_lat}) describes $2$ flavours $\Psi^T=( u,d)\times 16 $ doublers even in the $b\rightarrow 0$ limit. In fact, the Wilson-like term does not remove the doublers because it has dimension six. This makes no harm in this quenched study aimed at testing whether the mass generation mechanism occurs at all.
For further unquenched studies domain-wall \cite{DW} or overlap fermion \cite{Neuberger} will be required.


One can check that the action~(\ref{L_lat}) is invariant under global 
$\chi_L \otimes \chi_R$ transformations (see eq.(\ref{CHIL}))
and the lattice version of the discrete 
$P$, $T$, $C$ and $CPF_2$ symmetries. 
The discretization of the covariant derivatives in the Wilson-like terms of 
$D_{lat}$ (the ones with coefficient $\rho$) is chosen so that the lattice action 
$S_{lat}$ is exactly invariant under the "spectrum doubling symmetry"~\cite{Montvay}.  
\begin{eqnarray}
\Psi(x) \to \Psi'(x)=e^{-ix\cdot\pi_H}M_H\Psi(x) \quad\quad \overline\Psi(x)\to \overline\Psi'(x)=\overline\Psi(x)M_H^\dagger e^{ix\cdot\pi_H}
\end{eqnarray}
where $H$ is an ordered set of four-vector indices $H\equiv\{\mu_1,...,\mu_h\},\,(\mu_1<\mu_2<...<\mu_h)$. For $0\leq h \leq 4$ there are 16 four-vector $\pi_H$ with $\pi_{H,\mu}=\pi$ if $\mu\in H$ otherwise $\pi_{H,\mu}=0$ and 16 matrices
$M_H\equiv (i\gamma_5\gamma_{\mu_1})...(i\gamma_5\gamma_{\mu_h})$.
The fact that only symmetric derivatives $\widetilde \nabla$ appear in the 
Wilson-like actions terms and the consequent "spectrum doubling symmetry" 
guarantee that \\
{\bf a)} at tree level the Wilson-like terms contribute only O($b^2$) 
effects as it is clear by noting e.g. \\ 
\begin{equation}
 -b^2 \sum_x e^{ipx} \widetilde\nabla_\mu \widetilde\nabla_\mu  \Psi(x) 
\vert_{ p = (0,b^{-1}\pi + \epsilon,0,0) \equiv \bar p } = \sin^2(\pi + b\epsilon)
\tilde\Psi(p)\vert_{ p = \bar p } = (b^2\epsilon^2 + O(b^4\epsilon^4)) 
\tilde\Psi(\bar p) \, , 
\end{equation}
\begin{equation}
\sum_y e^{ipy} \gamma_\mu \widetilde\nabla_\mu  \Psi(y) 
\vert_{ p = (0,b^{-1}\pi + \epsilon,0,0) \equiv \bar p } = 
-i b^{-1} \gamma_\mu \sin(bp_\mu)\tilde\Psi(p)\vert_{ p = \bar p }
= i\gamma_2 (\epsilon + O(b^2\epsilon^3))\tilde\Psi(\bar p) \; ;
\end{equation}
{\bf b)} beyond tree level, as far as removal of UV divergencies is concerned,
the situation is like it would be in the $\rho=0$ case: only renormalization 
of the fermion kinetic term ($\bar\Psi \widetilde \nabla \Psi$) and Yukawa
coupling ($\eta$) is needed, besides the usual renormalization of 
gauge and scalar fields and parameters.

This implies in particular that $\eta_{cr}$, the critical value of $\eta$,  
is well defined (even in the presence of fermion doubling) and independent
from the subtracted scalar squared mass $\mu_{sub}^2$ (thus equal for 
the Wigner phase and the Nambu-Goldstone phase).

\section{Lattice procedure and correlators}

In order to confirm (or falsify) the mass generation mechanism we need to study
the renormalized $\tilde\chi_L$--WTIs (eq.~(\ref{ren_WTI})) and hence to evaluate
at least two-point correlators of the form
\begin{equation}
 \partial_\mu\langle  \widetilde J^{L,i}_\mu(x) {\cal O}^i(z) \rangle 
\quad {\rm and} \quad 
\langle B^{L ,i}_{Yuk}(x) {\cal O}^i(z) \rangle 
\,\,\quad  x\neq z\;, \quad
\end{equation}
where $B^{L ,i}_{Yuk}$ stands for the variation of the Yukawa term under 
$\tilde\chi_L$ (see eq.~(\ref{BYUK})) and $\widetilde J^{L,i}_\mu$ is the lattice
version of the current (\ref{JCLT}) given the action (\ref{L_lat}). The local
operator ${\cal O}^i$ is taken conveniently so as to avoid vanishing correlators. 

Our procedure starts in the Wigner phase by choosing reasonable values of 
$g_s^2$ (hence $b\Lambda_s$), $\rho$ and $\lambda_0$ and looking for the 
(critical) value of $\eta$ where
 \begin{equation}
 \langle \partial_\mu \widetilde J^{L,i}_\mu(x) {\cal O}^i(z) \rangle\big{|}_{\eta_{cr}}=0  \, ,\quad 
\,\,  \mu_{sub}^2>0\nn
 \end{equation} 
As next step we move to the Nambu--Goldstone phase keeping
$\eta$ fixed at its critical value, $\eta_{cr} = \eta_{cr}(g_s^2,\rho, \lambda_0)$
and we check whether  
 \begin{equation}
  \frac{ \langle \partial_\mu \widetilde J^{L,i}_\mu(x) {\cal O}^i(z) \rangle\big{|}_{\eta_{cr}}}
  {\langle B^{L ,i}_{Yuk}(x) {\cal O}^i(z) \rangle\big{|}_{\eta_{cr}}}
  = {\mbox{O}}(C_1{\Lambda_s}) {\neq 0} \, ,\quad   \mu_{sub}^2<0  \, .\label{CONDI}
 \end{equation} 
In the context of the mechanism under study, the dimensionless coefficient $C_1$
should become independent of the scalar vev $v \simeq | \mu_r^2 \lambda_r|$ when
$\Lambda_s^2 \ll v^2 \ll b^{-2}$. Finally one has to check the result for $C_1$
as the continuum limit ($b \to 0$) is taken at some fixed renormalization condition.  

Since fermions are quenched, scalar and gauge field configurations can be generated
independently from each other.
As customary, we choose ${\cal O}^i=B^{L ,i}_{Yuk}$ and in order to reduce statistical
errors we study the ratio of zero three-momentum correlators
 \begin{equation}
 R_L(x_0)=\frac{\sum_{\vec x} \partial_\mu\langle  \widetilde J^{L,i}_\mu(\vec x,x_0) B_{Yuk}^{L,i}(\vec z,z_0) \rangle}{\sum_{\vec x} \langle B_{Yuk}^{L,i}(\vec x,x_0) B_{Yuk}^{L,i}(\vec z,z_0) \rangle} \quad x_0\neq z_0\label{RATIO}.
  \end{equation}

\subsection{Technical remarks}

In a numerical simulation on a finite lattice the scalar v.e.v.\ is always zero, even 
if $\mu_{sub}^2 <0$. Hence an `axial fixing'' of the global $\chi_L\times\chi_R$ 
symmetry~\cite{scalar} is carried out in order to get $\langle\Phi\rangle = v >0$ 
in the Nambu-Goldstone phase. In this phase an IR cut-off to correlators (and a 
non-zero lowest eigenvalue for the Dirac matrices to be inverted) will be provided
by the scalar v.e.v.\ if $\eta \neq \eta_{cr}$ and possibly (even at $\eta =\eta_{cr}$)
by the non-perturbatively generated fermion mass. In the Wigner phase however (we have
checked that) this is not the case and an external IR cutoff must be introduced in
the computations to determine $\eta_{cr}$~\footnote{This is even more necessary in the
quenched approximation, when obviously there is no fermion determinant suppression for 
"gauge-scalar" configurations supporting zero modes of $D_{lat}[U,\Phi]$.}. 
One simple way out is to compute all correlators by approximating $D_{lat}^{-1}$ with
$D_{lat}^\dagger ( D_{lat} D_{lat}^\dagger + M_0^2)^{-1}$ for a number of small values
of $M_0^2$ and then take the limit $M_0^2 \to 0$ in the ratio~(\ref{RATIO}), which
allows to determine $\eta_{cr}$ and is hopefully smoothly depending on $M_0^2$.
Another possible approach is to add a term $\sum_x \bar \Psi(x) m \Psi(x)$ to the
action~(\ref{L_lat}), which provides the desired IR cut-off (since $(D_{lat}+m)^{-1}$ 
now enters in correlators) while breaking only in a {\em soft} way the otherwise
exact $\chi_L \otimes \chi_R$ symmetry of the model and not affecting $\eta_{cr}$. 

%
%
\subsection{First evidence of a signal}

\begin{figure}
\begin{center}
\includegraphics[scale=.75]{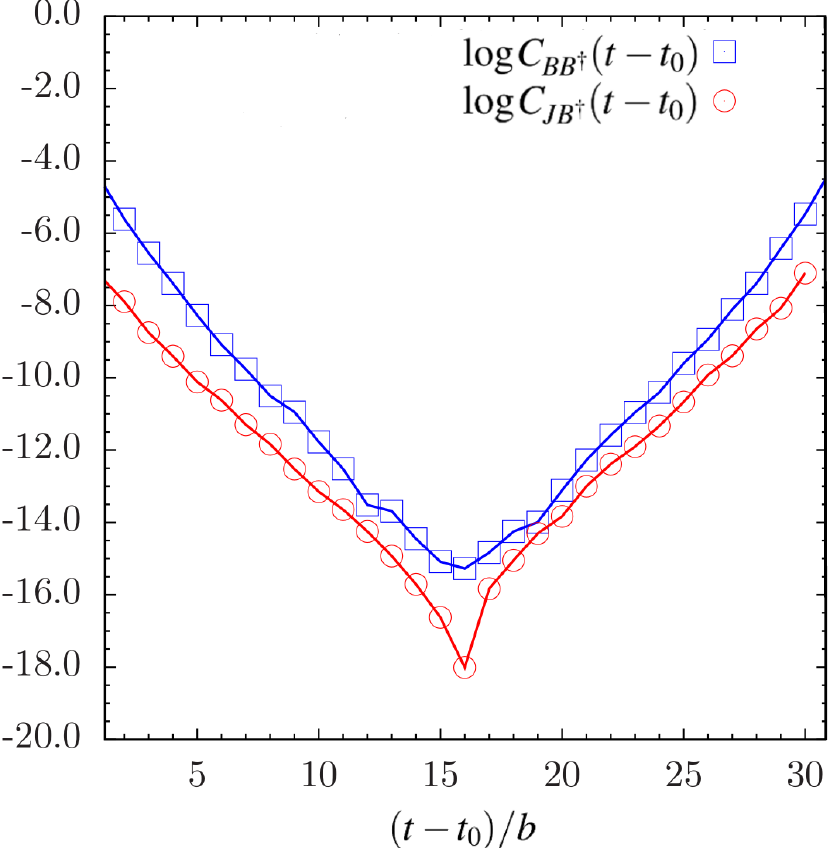}
\caption{Time dependence of $C_{BB^\dagger}$ and $C_{JB^\dagger}$ 
on $16^3 \! \times \! 32$ lattice: reliable errorbars not yet available.  
}
\label{fig:C_DD}
\end{center}

\vspace{-1cm}
\end{figure}

We have started a first exploration of the signal for the correlators 
entering~(\ref{RATIO}) in the Wigner phase. We choose bare parameters 
such that $b\Lambda_s \sim 0.1$ ($\beta=6/g_0^2=5.85$),
$b^2\mu_{sub}^2 \sim 0.072$, $\lambda_0 \sim 0.592$, 
$\rho=1$ and $\eta=0.2$ on a lattice with $16^3 \times 32$ sites. 
To have an IR-cutoff in place 
we set $b^2M_0^2 = 0.0005$ (to be varied later). 
As shown in fig.~\ref{fig:C_DD}, we evaluate the correlators 
$ C_{BB^\dagger}(t-t_0)=
\sum_{{\bf x}} \langle B_{Yuk}({\bf x},t)B_{Yuk}^\dagger(x_0,t_0) \rangle$ 
and 
$C_{JB^\dagger}(t-t_0)=
\sum_{{\bf x}} \langle \tilde J^{L,i}_0({\bf x},t)B_{Yuk}^\dagger(x_0,t_0) \rangle$, 
where 
\begin{eqnarray}
B_{Yuk}({\bf x},t)= \overline \Psi({\bf x},t) \frac{\tau^i}{2}\Phi({\bf x},t)\left( \frac{1+\gamma_5}{2}\right) \Psi({\bf x},t)-
\overline \Psi({\bf x},t) \Phi^\dagger({\bf x},t)\frac{\tau^i}{2}\left( \frac{1-\gamma_5}{2}\right) \Psi({\bf x},t)
 \label{BYUK} \end{eqnarray}
as functions of the Euclidean time separation $t-t_0$ and get a signal
while the correlator magnitude varies by more than 10 orders of magnitude. 
Note that in this hyper-preliminary example 
our statistics is very limited: just $8$ different scalar configurations times $9$ 
gauge configurations (well decorrelated from each other).
In order to improve the signal the action~(\ref{L_lat}) is modified by replacing
(only) in the term $\sum_x \bar \Psi(x) D_{lat}[U,\Phi] \Psi(x)$ the scalar field
$\Phi(x)$ with its average over the $\Phi$--values at the sites corresponding to
the 16 vertices of the hypercube of side $2b$ centered in $x$. Moreover, we also 
carry out a spatial smearing of the resulting $\Phi$ field entering in $B_{Yuk}$.

\end{document}